\documentclass{article}
\usepackage{spconf,amsmath,graphicx,hyperref}

\usepackage{multirow}
\usepackage{booktabs}
\usepackage{bbding}

\usepackage{dblfloatfix}

\title{AudioRAG+: Feedback-driven Retrieval-augmented Audio Generation with Large Audio Language Models}
\name{
\begin{tabular}{c}
Junqi Zhao\textsuperscript{1,2},
Chenxing Li\textsuperscript{2},
Jinzheng Zhao\textsuperscript{1,2},
Rilin Chen\textsuperscript{2},\\
Dong Yu\textsuperscript{3},
Mark D. Plumbley\textsuperscript{1},
Wenwu Wang\textsuperscript{1}
\end{tabular}
}

\address{$^{1}$ Centre for Vision, Speech and Signal Processing (CVSSP), University of Surrey, UK \\
         $^{2}$ Tencent AI Lab, Beijing, China \\
         $^{3}$ Tencent AI Lab, Seattle, USA}
         
\begin{document}
\sloppy

%
\maketitle
\begin{abstract}
We propose a general feedback-driven retrieval-augmented generation (RAG) approach that leverages Large Audio Language Models (LALMs) to address the missing or imperfect synthesis of specific sound events in text-to-audio (TTA) generation. Unlike previous RAG-based TTA methods that typically train specialized models from scratch, we utilize LALMs to analyze audio generation outputs, retrieve concepts that pre-trained models struggle to generate from an external database, and incorporate the retrieved information into the generation process. Experimental results show that our method not only enhances the ability of LALMs to identify missing sound events but also delivers improvements across different models, outperforming existing RAG-specialized approaches.

\end{abstract}
\section{Introduction}
\label{sec:intro}

Text-to-audio (TTA) generation refers to the task of generating audio content conditioned on natural language descriptions. This field has been transformed by recent advances in generative models, especially diffusion-based \cite{liu2023audioldm, ghosal2023text, hung2024tangoflux, zhao2025audioturbo} and language model-based approaches \cite{kreukaudiogen, yang2023uniaudio, yanggenerative}. 

These generative models require extensive training data and computational resources. Due to their reliance on the training data, they struggle to generate audio categories that are rare or unseen in the dataset \cite{yuan2024retrieval}, along with user-specified audio content. This limitation substantially hinders their generative performance.

To address the class-imbalance problem in the training data of TTA models, Re-AudioLDM \cite{yuan2024retrieval} builds upon the AudioLDM \cite{liu2023audioldm} model and retrieves relevant audio-text pairs from the database as supplementary information, thereby improving the modeling of certain rare sound events during training. Audiobox TTA-RAG \cite{yang2024audiobox}, a recent retrieval-augmented TTA method based on Audiobox \cite{vyas2023audiobox}, extends the flow-matching model by augmenting the conditioning input with both text prompts and retrieved audio samples, and eliminates the need for labels in the external audio data source. However, these approaches necessitate training a dedicated RAG-assisted Latent Diffusion Model (LDM) or Flow Matching (FM) model from scratch. This significantly increases the training cost, and whenever a new concept must be generated, the model’s architecture upgraded, or the database updated, the model needs to be retrained.



In contrast to prior RAG-specialized TTA models, we propose a novel feedback-guided retrieval augmentation approach built upon a Large Audio Language Model (LALM) to address the limitations of current TTA models, particularly their suboptimal generation performance and the omission of certain sound events. Specifically, we first use the fine-tuned LALM \cite{xu2025qwen2} to evaluate the audio generated by the pre-trained TTA model, assessing whether it aligns with the text prompt and identifying any missing or suboptimal sound events. Next, for audio concepts that the model struggles to generate effectively, we retrieve relevant samples from an external database. Finally, we employ a decoupled cross-attention mechanism to integrate the retrieved audio as references into the pre-trained TTA model during inference, steering the model toward producing the desired audio outputs and thereby enhancing its overall generation efficacy.

We conducted experiments on different pre-trained TTA models, including both diffusion-based  \cite{liu2024audioldm} and flow-matching-based \cite{hung2024tangoflux} models. We will show that our method can be applied across all pre-trained TTA models, without the need to train a dedicated retrieval-augmented TTA model from scratch as required by approaches such as Re-AudioLDM or Audiobox TTA-RAG, thereby ensuring broad applicability. Our proposed method achieves substantial performance improvements across different TTA models, outperforming the state-of-the-art (SOTA) RAG-based TTA approaches by a significant margin.

\begin{figure*}[htb]
  \centering
  \includegraphics[width=16cm]{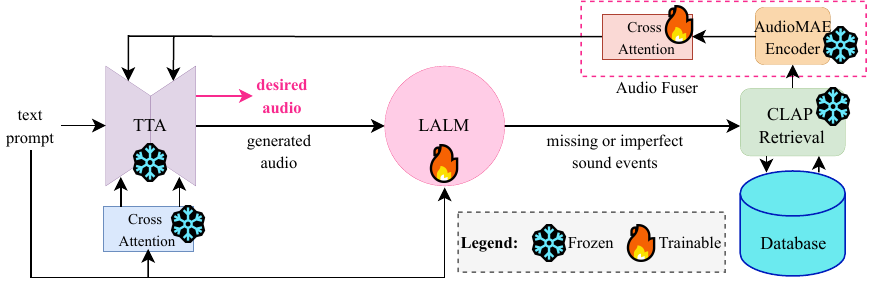} 
  \caption{The overview structure of our proposed method.}
  \label{fig:overview}
\end{figure*}

Our main contributions can be summarized as follows: 1) We introduce a feedback-driven RAG method based on an LALM to enhance the performance of pre-trained TTA models. 2) To improve the accuracy of the LALM in identifying missing sound events, we further fine-tune it using Low-Rank Adaptation (LoRA). 3) To tackle the challenges faced by current TTA models in complex multi-event audio scenes, we develop a lightweight audio fuser with decoupled cross-attention, which enhances audio events that are missing or poorly generated during inference.

\section{Releated Work}
\label{sec:format}

Retrieval-augmented generation, first introduced in natural language processing (NLP) \cite{lewis2020retrieval, gao2023retrieval}, has emerged as a promising approach to mitigate hallucination and outdated knowledge in large language models (LLMs). Subsequently, RAG has attracted increasing research attention in the field of image generation \cite{blattmann2022retrieval, sheyninknn, chenre, shalev2025imagerag}. For speech generation tasks, Xue et al. \cite{xue2024retrieval} introduced a Context-Aware Contrastive Language-Audio Pre-training model that captures style-related contextual features and leverages contextual information in the retrieval process to improve the selection of specialized audio prompts. AutoStyle-TTS \cite{luo2025autostyle} presents a text-to-speech framework leveraging RAG, which dynamically adapts speech styles to textual context, thereby enabling more natural and expressive speech synthesis. 

For the TTA generation task examined in this study, Re-AudioLDM \cite{yuan2024retrieval} and Audiobox TTA-RAG \cite{yang2024audiobox} attribute the performance limitations to class imbalance in the training datasets and therefore train a dedicated RAG-based model from scratch on top of a base TTA model.


\section{Proposed Method}
\label{sec:pagestyle}

As shown in Figure~\ref{fig:overview}, our proposed method consists of three main components: a finetuned LALM module that evaluates the outputs of pre-trained TTA models; a retrieval module that searches the database for relevant audio using missing or imperfect sound event captions generated by the LALM; and a TTA model enhanced with an audio fuser.

The workflow of our system proceeds as follows: we begin by augmenting the pretrained TTA model with a decoupled cross-attention module for audio, yielding an audio-enhanced TTA model. During the retrieval stage, the parameters of the audio-enhanced TTA model remain fixed. For the currently popular pre-trained TTA models, the input is the textual description of the target audio and the output is the corresponding audio clip. However, when faced with complex audio scenes or rare and even unseen audio types, the model may only generate a subset of the intended sound events, or the quality of certain generated events may be unsatisfactory. A fine-tuned LALM can interpret the audio outputs of the pre-trained TTA model and identify sound events that are either missing or poorly generated, thereby enabling targeted enhancement. This enhancement is achieved by retrieving the corresponding audio from an external database and feeding it back to the pre-trained TTA model through our designed audio fuser.

\subsection{Large Audio Language Model}
\label{ssec:subhead}

Qwen2.5-Omni \cite{xu2025qwen2} represents a SOTA LALM capable of simultaneously processing auditory and textual information and producing real-time outputs in both text and natural speech. It adopts a Thinker-Talker architecture.

Although Qwen2.5-Omni demonstrates strong performance across various audio understanding benchmarks, its effectiveness is limited in the missing or imperfect sound event recognition task required in this study. To address this issue, we fine-tune the Qwen2.5-Omni-7B model on our task-specific dataset using the LoRA approach. More concretely, we build a dataset for the task of missing sound event identification. Each instance consists of an audio sample for the LALM to interpret and a text query: “What sound events are missing from the audio compared to \texttt{<target prompt>}?” The LALM is required to output the sound events that are not present in the audio but appear in the target prompt.
 
\subsection{Retrieval Module}
\label{ssec:subhead}
Similar to Re-AudioLDM and Audiobox TTA-RAG, we employ the Contrastive Language–Audio Pretraining (CLAP) model as the retrieval module to query the database for audio samples based on the text descriptions of missing or imperfect sound events. Since our approach adopts text-to-audio retrieval, it does not require labeled external data sources, thereby supporting large-scale retrieval from in-the-wild and unlabeled audio datasets.

\subsection{Audio-Enhanced Text-to-Audio Generation}
\label{ssec:subhead}


The audio-enhanced TTA model consists of two components: an original pre-trained TTA model and an audio fuser. The audio fuser is designed to enable the pre-trained TTA model to generate audio conditioned on retrieved audio prompts. To allow the TTA model to capture fine-grained features from the audio prompt, we first employ an Audio Masked Autoencoder (MAE) \cite{huang2022masked} to extract representations from the input audio. We then introduce an adapted module based on a decoupled cross-attention mechanism \cite{ye2023ip}, which separates the cross-attention layers for audio and text, in order to inject these features into the internal layers of the pre-trained TTA model.

In the original pre-trained TTA, the text encoder features $c_\mathrm{t}$ are injected into a diffusion model through cross-attention layers:

\begin{equation}
z_{\mathrm{t}} = \text{Attention}\big(zW_{q}, \; c_\mathrm{t} W_{k}, \; c_\mathrm{t} W_{v}\big).
\end{equation}
Here, $z$ denotes the internal feature of the diffusion model; 
$Q = zW_q$, $K = c_\mathrm{t}W_k$, and $V = c_\mathrm{t}W_v$ correspond to the query, key, 
and value in the attention operation, respectively, where 
$W_q$, $W_k$, and $W_v$ are the weight matrices of the linear 
projection layers.


Instead of inserting audio features by concatenating them with text features and feeding the result into the cross-attention layers, we add an additional cross-attention layer to each cross-attention layer in the original TTA model to integrate audio features. Given the AudioMAE features $c_\mathrm{a}$, the output of the new cross-attention is computed as follows:
\begin{equation}
    z_\mathrm{a} = \text{Attention}\big(zW_{q}, \; c_\mathrm{a} W_{k}', \; c_\mathrm{a} W_{v}'\big),
\end{equation}
where $W'_k$ and $W'_v$ are the newly added weight matrices. Then, we add the output of the audio cross-attention to that of the text cross-attention via:
\begin{equation}
  z^{\mathrm{new}} = z_\mathrm{t} + \lambda z_\mathrm{a}
\end{equation}
where $\lambda$ is a weight factor that controls the strength of the retrieved audio.

In the training phase of the audio-enhanced TTA model, 
the parameters of the original TTA model are frozen, while only $W'_k$ and $W'_v$ are updated, 
enabling parameter-efficient fine-tuning and reducing training overhead.


\section{Experiments}
\label{sec:typestyle}

\subsection{Datasets}
\label{datasets}


We constructed a training dataset for the missing sound event identification task using the AudioCaps (AC) training dataset and the AudioSet (AS) balanced training dataset, comprising a total of 45,222 samples. Similarly, we construct the test set using the AudioCaps test set and the AudioSet evaluation set.


For training the audio-enhanced TTA model, we only train the cross-attention layers for audio using the AC training set. The external database consists of the AS balanced subset and the Freesound dataset. For RAG evaluation, we employ the AC test set along with the RiTTA \cite{he2024ritta} Count test set.

\begin{table}[t]
\centering
\small
\begin{tabular}{lcc}
\toprule
\textbf{Model} & \textbf{BERTScore (\%)} & \textbf{SimCSE (\%)} \\
\midrule
Ground Truth            & 100.0          & 100.0 \\
Gemini 2.5 Pro          & 80.2           & 89.1 \\
Qwen2.5-Omni-7B         & 53.8           & 73.8 \\
Qwen2.5-Omni-7B (SFT)   & \textbf{93.3}  & \textbf{92.6} \\
\bottomrule
\end{tabular}
\caption{Comparison of different models for missing sound event identification.}
\label{tab:missing_sei}
\end{table}

\begin{table*}[t]
\centering
\small
\begin{tabular}{cccccccc}
\toprule
 Model  & Dataset & Retrieval Info. & Database \& Retrieval No.  & KL $\downarrow$ &  FD $\downarrow$ &  IS $\uparrow$ & $\operatorname{CLAP}$ (\%) $\uparrow$ \\


\midrule
Re-AudioLDM-L &AudioCaps&Audio \& Text&  AC \textrightarrow 10 &   $\mathbf{1.20}$ & $-$   &  $7.39$  & $37.12$ \\

\midrule
Audiobox TTA-RAG &AudioCaps&Audio& AC \textrightarrow 3 &   $1.44$ & $-$ &  $8.40$  & $37.37$ \\

\midrule
AudioLDM2 &AC+AS+6 others&\XSolidBrush &  \XSolidBrush &   $1.59$ & $33.2$ &  $7.40$  & $45.20$ \\

\midrule
AudioLDM2-RAG (ours) &AC+AS+6 others&Audio &  AS \textrightarrow 1 &   $1.55$ & $30.6$ &  $8.49$  & $46.22$ \\

\midrule
TangoFlux &AC+1 other&\XSolidBrush &  \XSolidBrush &   $1.21$ & $19.23$ &  $12.60$  & $\mathbf{58.60}$ \\

\midrule
TangoFlux-RAG (ours) &AC+1 other&Audio &  AS \textrightarrow 1 &   $\mathbf{1.20}$ & $\mathbf{18.98}$ &  $\mathbf{12.81}$  & $\mathbf{58.60}$ \\

\bottomrule
\end{tabular}
\caption{Performance comparison of different models on the AudioCaps test set (ID).}
\label{tab:ID}
\end{table*}

\subsection{Experimental Setup}
\label{ssec:exp_setup}
\textbf{Implementation Details.}
For Qwen2.5-Omni, we fine-tune the 7B version of the model with a LoRA rank of 8. The model is trained for 5 epochs with a batch size of 4.
As for the audio-enhanced TTA model, we employ two base models, AudioLDM2-Large \cite{liu2024audioldm} and TangoFlux \cite{hung2024tangoflux}. All parameters in the pre-trained TTA model and the AudioMAE are frozen, except for the decoupled audio cross-attention layers. Training is conducted for 20,000 steps with an effective batch size of 28. We adopt the AdamW optimizer with a fixed learning rate of $1 \times 10^{-4}$ and a weight decay of $1 \times 10^{-2}$. To enable classifier-free guidance at inference, we randomly drop audio and text conditions with a probability of 5\% during training.

\noindent
\textbf{Evaluation Metrics.}
Following Re-AudioLDM and Audiobox TTA-RAG, we evaluate the effectiveness of our proposed method using Fréchet Distance (FD), Kullback–Leibler (KL) divergence, Inception Score (IS), and CLAP score. Lower FD and KL scores reflect higher audio quality, while a higher IS score indicates better quality and greater diversity in the generated audio. The CLAP score measures the degree of semantic alignment between the generated audio and the given text.

In addition, to evaluate the accuracy of the LALM in the missing sound event identification task, we adopt BERTScore \cite{zhangbertscore} and SimCSE \cite{gao2021simcse} as evaluation metrics, as they are important measures of textual similarity in NLP.









\section{Results}
\label{sec:typestyle}

\subsection{Missing Sound Events Identification}
\label{missing_sdi}



In our experiments, we observed that Qwen2.5-Omni tends to generate hallucinations in the task of identifying missing sound events. To address this issue, we performed supervised fine-tuning (SFT) on our custom dataset. The experimental results of different LALMs on the test set are presented in Table~\ref{tab:missing_sei}. The results indicate that although Qwen2.5-Omni-7B achieves state-of-the-art performance on many common audio understanding tasks, its zero-shot performance on the missing sound event identification task remains unsatisfactory. With supervised fine-tuning, however, the performance of Qwen2.5-Omni-7B (SFT) improves substantially and even surpasses that of Gemini 2.5 Pro. This fine-tuning enables us to leverage LALMs to more effectively evaluate the outputs of pre-trained TTA models in identifying missing or low-quality sound events, thereby facilitating the retrieval of relevant audio from external databases to enhance generation performance.

\subsection{Main Results}
\label{main_resu}

The experiments are conducted on two test sets: the AudioCaps test set and the RiTTA Count test set, which represent in-distribution (ID) results and out-of-distribution (OOD) generalization results, respectively. We compare our performance against several state-of-the-art models, including Re-AudioLDM \cite{yuan2024retrieval}, Audiobox TTA-RAG \cite{yang2024audiobox}, AudioLDM2 \cite{liu2024audioldm}, and TangoFlux \cite{hung2024tangoflux}.

\subsubsection{RAG Results on AudioCaps}
\label{resu_AC}

The generation performance of different models on the AudioCaps dataset is presented in Table~\ref{tab:ID}. We use the AudioSet balanced subset as our database. When applying our proposed RAG method, both AudioLDM2-RAG and TangFlux-RAG exhibit notable performance improvements over their original TTA models. This demonstrates that our approach is highly effective across different base models. Moreover, it eliminates the need to train a dedicated RAG model from scratch, significantly reducing the training costs. In addition, our TangoFlux-RAG achieves the best performance compared with the baseline models across all evaluation metrics. The improvement over TangoFlux is not very large, which may be due to the fact that TangoFlux is currently the strongest TTA model, and that the training and test distributions are highly similar, which may introduce evaluation bias.

\subsubsection{RAG Results on RiTTA}













\begin{table}[t]
\centering
\small
\resizebox{\columnwidth}{!}{%
\begin{tabular}{cccccc}
\toprule
Model  & KL $\downarrow$ &  FD $\downarrow$ &  FAD $\downarrow$ &  IS $\uparrow$ & $\operatorname{CLAP}$ (\%) $\uparrow$ \\
\midrule
AudioLDM2 &   $2.81$ & $38.5$ &  $7.7$ &  $7.4$  & $29.0$ \\
\midrule
AudioLDM2-RAG (ours) &   $\mathbf{2.71}$&   $\mathbf{35.2}$&   $\mathbf{4.4}$ & $\mathbf{8.5}$ &  $\mathbf{34.2}$ \\
\hline\hline
TangoFlux &   $2.22$ & $46.8$ &  $7.3$ &  $7.0$  & $43.3$ \\
\midrule
TangoFlux-RAG (ours) &   $\mathbf{2.18}$&   $\mathbf{37.7}$&   $\mathbf{5.1}$ & $\mathbf{7.3}$ &  $\mathbf{43.7}$ \\
\bottomrule
\end{tabular}%
}
\caption{Performance comparison of different models on the RiTTA Count test set (OOD).}
\label{tab:OOD}
\end{table}

We selected the RiTTA Count test set for out-of-distribution evaluation, with the AudioSet balanced subset and the Freesound dataset as the database. Table~\ref{tab:OOD} reports the evaluation results of the base model and our RAG method. This experiment demonstrates that when the distributions of the training and test data differ, our method achieves better performance on the RiTTA test set than its base model. Overall, the poorer the model’s performance on a given metric, the greater the improvement obtained with our method.

\section{Conclusion}
\label{sec:conclu}

In this work, we present a general RAG method that can be applied to various pretrained TTA models and databases, eliminating the need to train a specialized RAG-based TTA model from scratch. To improve performance, we utilize an LALM to evaluate the outputs of the base model, and then introduce a lightweight audio fuser to incorporate the retrieved audio. Experiments show that our method delivers consistent performance improvements in both in-domain and out-of-domain settings, outperforming dedicated RAG models.

\ninept
\bibliographystyle{IEEEtran}
\bibliography{strings,refs}

\end{document}